\documentclass[3p,twocolumn]{elsarticle}
\usepackage{hhline}
\usepackage{fancyvrb}
\usepackage{amsmath}

\newcommand{\be}{\begin{equation}}
\newcommand{\ee}{\end{equation}}
\newcommand{\bea}{\begin{eqnarray}}
\newcommand{\eea}{\end{eqnarray}}
\newcommand{\p}{\partial}
\newcommand{\vvect}{\mathbf{v}}

\newcommand{\mynabla}{\pmb{\nabla}}

\begin{document}

\title{Hydrodynamics of evaporating sessile drops}

\author{L.Yu. Barash}
\ead{barash@itp.ac.ru}
\author{L.N. Shchur}
\address{Landau Institute for Theoretical Physics, 142432
Chernogolovka, Russia}

\begin{abstract}
Several dynamical stages of the Marangoni convection
of an evaporating sessile drop are obtained.
We jointly take into account the
hydrodynamics of an evaporating sessile drop,
effects of the thermal conduction in the drop and
the diffusion of vapor in air.
The stages are characterized by different number
of vortices in the drop and the spatial location of vortices.
During the early stage the array of vortices arises near
a surface of the drop and induces a non-monotonic spatial
distribution of the temperature over the drop surface.
The number of near-surface vortices in the drop
is controlled by the Marangoni cell size, which is calculated
similar to that given by Pearson for flat fluid layers.
The number of vortices quickly decreases with time,
resulting in three bulk vortices in the intermediate stage.
The vortex structure finally evolves into the single convection vortex
in the drop, existing during about $1/2$ of the evaporation time.
\end{abstract}

\maketitle

\section{Introduction}

Evaporation of a drop in an ambient gas was considered since Maxwell
time mainly as diffusion of vapor from a near-surface
layer~\cite{Maxwell,Langmuir1918,Fuchs}. Classical quasistationary
theory of evaporation of a drop does not include effects of fluid
dynamics in the drop and only partially takes into account basic
effects of heat transfer. Recently a number of new applications has
stimulated much attention to the problem again. This is
associated, for example, with preparing ultra-clean
surfaces~\cite{Leenaars90,Marra91,Huethorst91,Matar01}, protein
crystallography~\cite{Denkov,Dimitrov}, the studies of DNA stretching
behavior and DNA mapping methods~\cite{Jing,HuLarsonDNA,Hsieh},
developing methods for jet ink printing~\cite{Park,Jong,Lim}, and
with other fields (see, for example,~\cite{Frohn}). Of particular
interest is the process of evaporation of the drop containing the
colloidal suspension~\cite{Lin1,Lin2,Bigioni3}. One of typical
experimental situations is an observation of an evaporating drop lying
on a substrate, where the contact line of the drop is pinned to the
edge of the substrate. Ordered structures of nanoparticles can arise
on the drop surface during the evaporation process. After the drop
dries out the ordered structures are left also on the substrate.  An
important example is the self-assembly of long-range-ordered
nanocrystal superlattice monolayer~\cite{Lin1,Lin2,Bigioni3}.  Other
example, where depinning of the contact line is revealed, is the
effect of evaporative contact line deposition, the so-called
coffee-ring effect~\cite{Deegan97,Deegan,Govor,HuLarsonCoffee,
Popov1,Popov2}. These aspects altogether resulted in the interest to
the problem and respective significant activity of experimentalists
and theorists during the past decade.

For theoretical description of problems mentioned above, the
classical theory of drop evaporation, which was used up to recent
years, turned out to be insufficient, too simplificated and not
including a number of important processes. For example, role of
pinned contact line during the evaporation process, and fluid
dynamics effects within the evaporating drops of capillary size, were
analysed and new important aspects were found. As a result, a
development in experimental studies and progress in understanding
of evaporation process have been achieved during past decade~\cite{Deegan97,
Deegan,Popov1,HuLarsonEvap,HuLarsonMarangoni,Girard,Ristenpart}.

It was found, in particular, that the evaporating flux density is
inhomogeneous along the surface and has an integrated singularity on
approach to the pinned contact line~\cite{Deegan97,Deegan}.  The
resulting inhomogeneous mass flow can strongly modify the temperature
distribution over the drop surface and, hence, the Marangoni forces
associated with the temperature dependent surface tension.  The
convection inside a droplet~\cite{HuLarsonMarangoni,Girard,
Sefiane1,Sefiane2,Zhang,
Davis1,Davis2,Davis3,Davis4,Lozinsky,Niazmand,Rednikov,Savino,Xu}
appears to be quite different from the classical Marangoni convection
in the systems with a simple flat geometry~\cite{Benard,Pearson}.
Thermal conductivity of the substrate can also influence the
formation of flows within a liquid drop since it is the magnitude of
the conductivity which determines the sign of the tangential
component of the temperature gradient at the surface close to the
contact line and, therefore, the direction of the
convection~\cite{Ristenpart}.

The observation of the distinct stages of the evaporation
process~\cite{PicknettBexon,Shanahan1,Shanahan2} has revealed that
the longest and dominating regime of the evaporation process is the
constant contact area mode, where the contact line is pinned.  On a
later stage the contact line gets depinned and the different regime,
the constant contact angle mode switches on. Finally, the drying mode
follows, in which the height, the contact area and the contact angle
rapidly decrease with time.

Recently, evolved numerical calculations have been carried out
for providing simultaneous
descriptions of the fluid dynamics, the vapor diffusion and the
spatial temperature distribution in an axially symmetrical evaporating sessile drop. 
The corresponding quasistationary problem
was studied in~\cite{HuLarsonMarangoni,Girard}.  The
dynamical description of the liquid drop evaporation was developed
in~\cite{Barash1}. Taking into account quantitatively
the important components of
the evaporation process, we have obtained, in particular, 
the time evolution of the
temperature and the fluid convection in the drop.

\section{Dynamics of vortex structure}

\begin{figure*}[htb]
\caption{(Color online.)
Left panel: The velocity field at $t=0.16$ {\rm s}.
Right panel:
Temperature (solid line) and
velocitiy (dashed line) along the drop surface as functions
of the radius at $t=0.16$ {\rm s}.
The velocity changes its sign at crossing points with
the horizontal axis.
}
\label{karman}
\centering
\includegraphics[width=0.68\columnwidth]{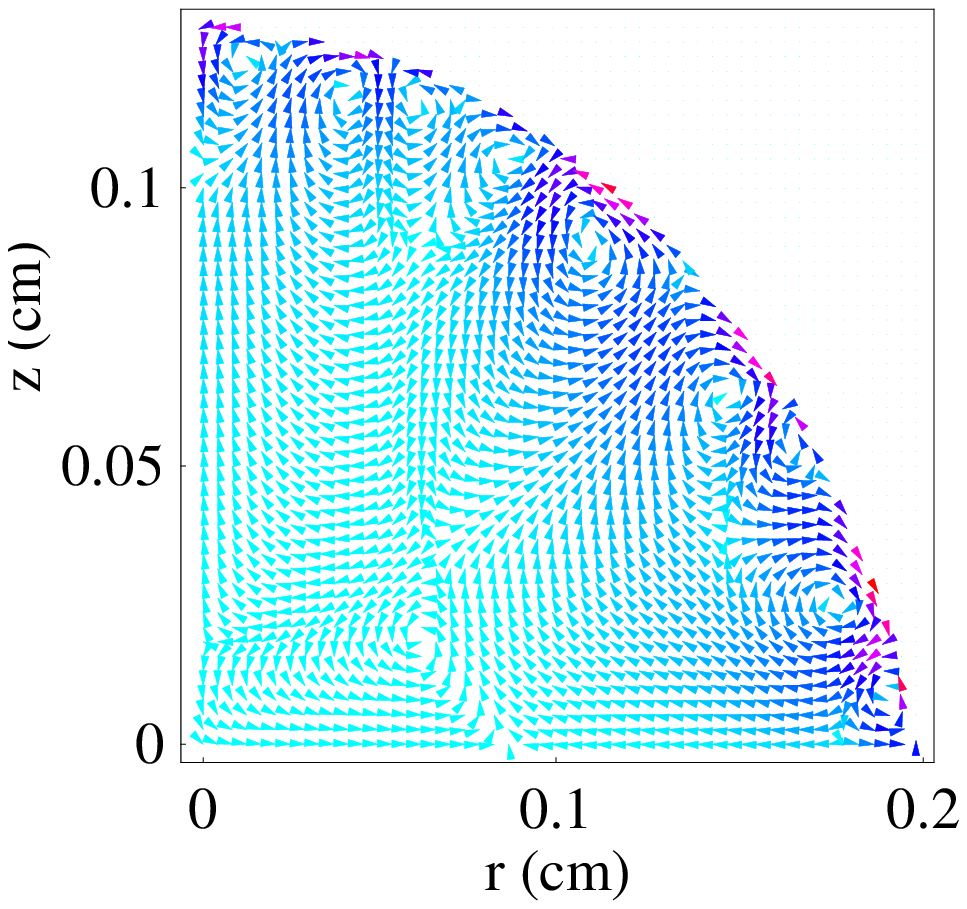}
\hspace{1.5cm}
\includegraphics[width=1.15\columnwidth]{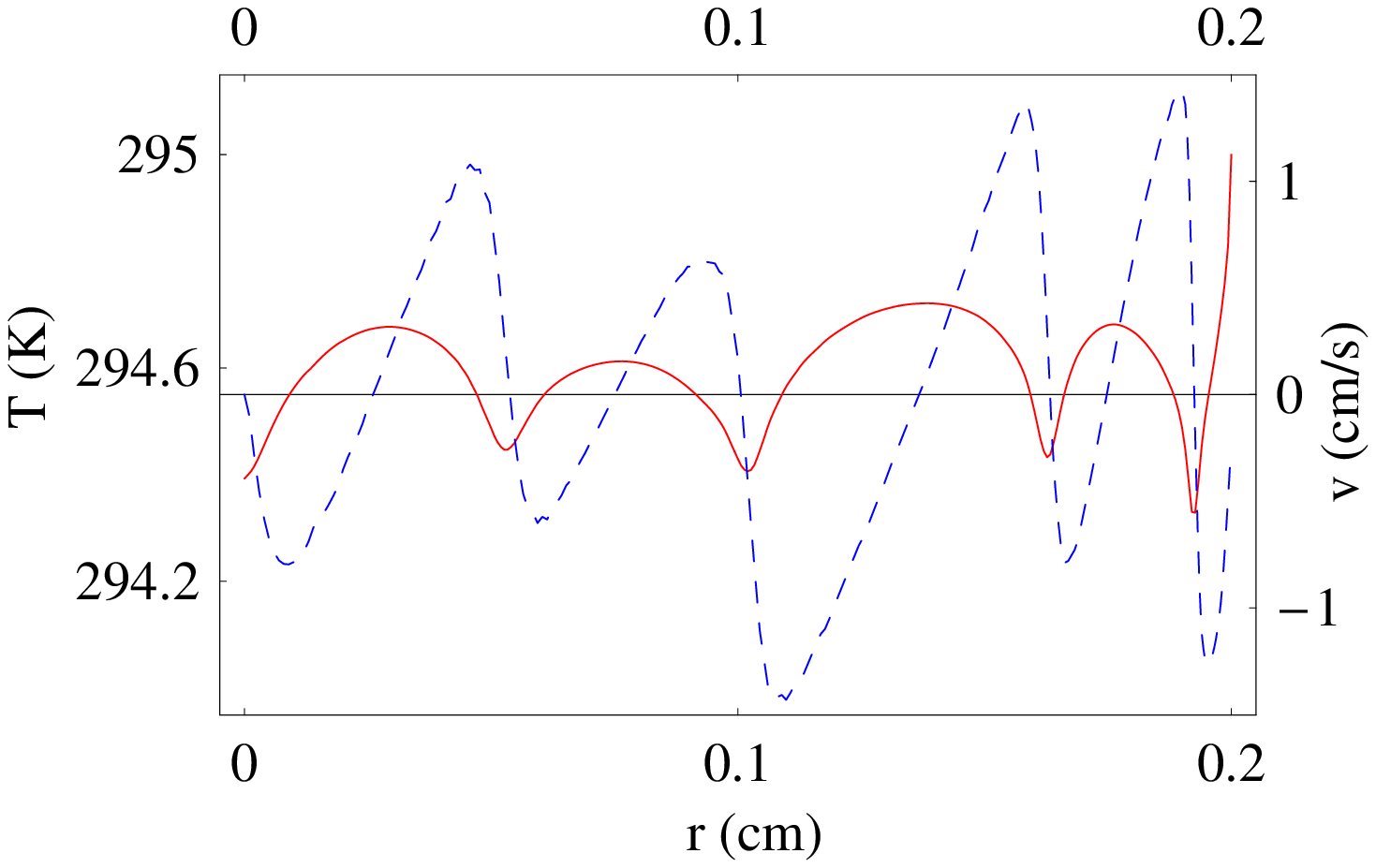}
\end{figure*}

Our calculations demonstrate the presence of several
characteristic stages of the thermocapillary convection.
During the early regime of the dynamics of the Marangoni convection,
for various liquids and drop sizes, the vortices arise
near the surface of the drop.
For the toluene drop,
this regime quickly arises and
evolves up to $t\approx 0.3$ s.
This is quite a short time period for the toluene drop
as compared with the total evaporation time $\approx 550$ s,
but it admits an experimental study.
The vortices grow,
the number of vortices decreases, and eventually
they evolve into the bulk of the drop. This can be seen
in Fig.~\ref{karman}
where the vortex structure contains
four pairs of near-surface vortices and a
corner vortex, and the temperature
displays just four humps at the surface.
This number of vortices in the drop
is controlled by the Marangoni cell size, which is
calculated similar to that given by Pearson
for flat fluid layers.
The existence of near-surface vortices and the associated
humps in the profile of the surface temperature,
become more pronounced with
the decrease of the viscosity of the liquid.
There are no near-surface vortices when the viscosity
increases more than in four times as compared with
the toluene drop.
Near-surface vortices were observed experimentally
in studying evaporating sessile drops of FC-72
on a copper substrate~\cite{Sefiane1,Sefiane2}.

As seen in Fig.~\ref{karman}, the extrema
of the surface temperature correspond to the
change of sign of the tangential component of
velocities at the surface. The reason for
this behavior is that the fluid flow moves
from the higher to the lower temperature regions
of the surface, because
the surface tension decreases with increasing the temperature.
The flows result in a redistribution of the temperature
due to the convective heat transfer.
The number of the surface temperature humps and
the number of the near-surface vortices decrease during their evolution.

If the thermal conductivity of a substrate is large compared to
that of the liquid, then the temperature can be maintained
practically constant at the substrate-fluid interface.
This is the case, in particular, for the silicon nitride substrates
used in experiments~\cite{Lin1,Lin2,Bigioni3}. The silicon nitride
is a material with high thermal conductivity, approximately in three
orders larger than for the toluene. For this reason,
the boundary condition for the temperature distribution
at the substrate can be reduced to the constant temperature.
Heat transfer between the substrate and the drop
plays an important role in establishing the temperature
profile in the drop. High substrate conductivity also excludes
a possibility for
the reversal of the Marangoni flows~\cite{Ristenpart},
taking place for substrates with relatively small thermal conductivity.

The initial conditions chosen when generating Fig.~\ref{karman}
were: zero velocity values,
constant room temperature, and absence of vapor.
For a description of a real experiment they have to be slightly modified to include the weak stochastic
distribution of the surface temperature and velocities.
We have carried out such calculations
using a small-scale stochastic initial conditions.
This has changed the particular behavior of the
fluid dynamics only on the initial stage of the process,
where a large number of small-scale surface vortices
arise. Then the surface vortex structure quickly evolves into
exactly the same one as we obtained for the basic
initial conditions. This result demonstrates the
generic character of the near-surface vortex regime
in Fig.~\ref{karman} at the early
stage of the formation of the Marangoni convection.

An initial velocity field within the sessile drop
can also be strongly disturbed right after the drop has fallen
down on the substrate, or for some other reason.
We model such a situation by choosing random initial
conditions for the bulk velocity field, which are
in agreement with the continuity equation for the incompressible
fluid.
We find that strong disturbances of the bulk
velocities can noticeably modify the initial
stage of the drop dynamics, but they do not
modify its main stage. For the initial random
velocities of the order of $5$ \rm{cm/s}
(which well exceeds the typical velocity in the vortex
$1$~cm/s),
the difference between the dynamics
of the disturbed and resting drops disappears
after $t\approx 0.5$ \rm{s}.
This verifies the stability of the large-scale drop dynamics
with respect to disturbances of the initial temperature
and velocity fields.

\begin{figure*}[bht]
\caption{(Color online.) Left panel: The velocity distribution
at $t=0.5$ s. The stage of drop dynamics with three vortices
takes place from $t\approx 0.45$ s to $t\approx 2.0$ s.
Right panel: The velocity distribution at $t=30$ s.
A distribution with a single vortex takes place from
$t\approx 2.0$ s to $t\approx 250$ s.
} \label{vect2}
\hspace{0.9cm}
\includegraphics[width=0.65\columnwidth]{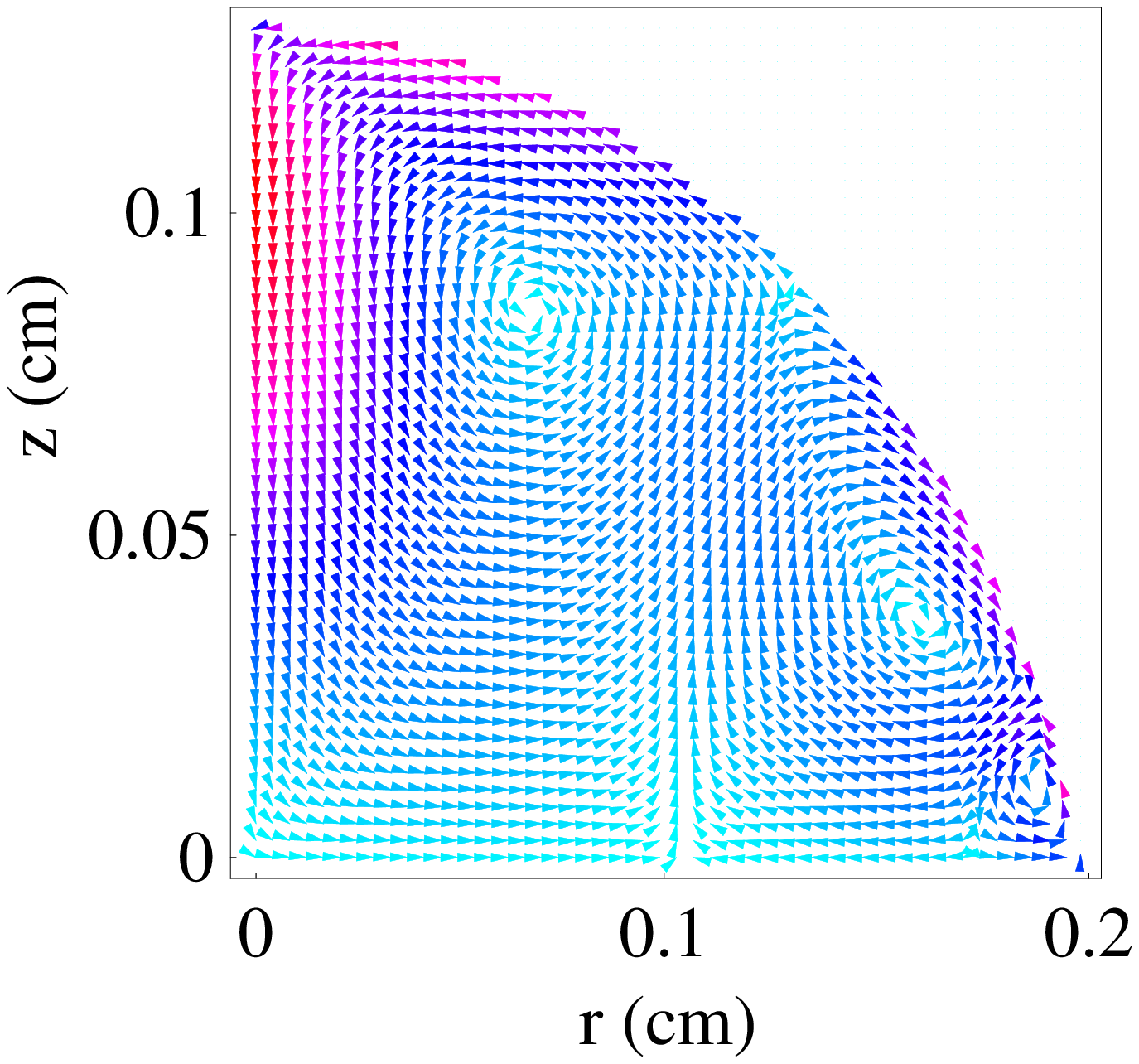}
\hspace{2.2cm}
\includegraphics[width=0.68\columnwidth]{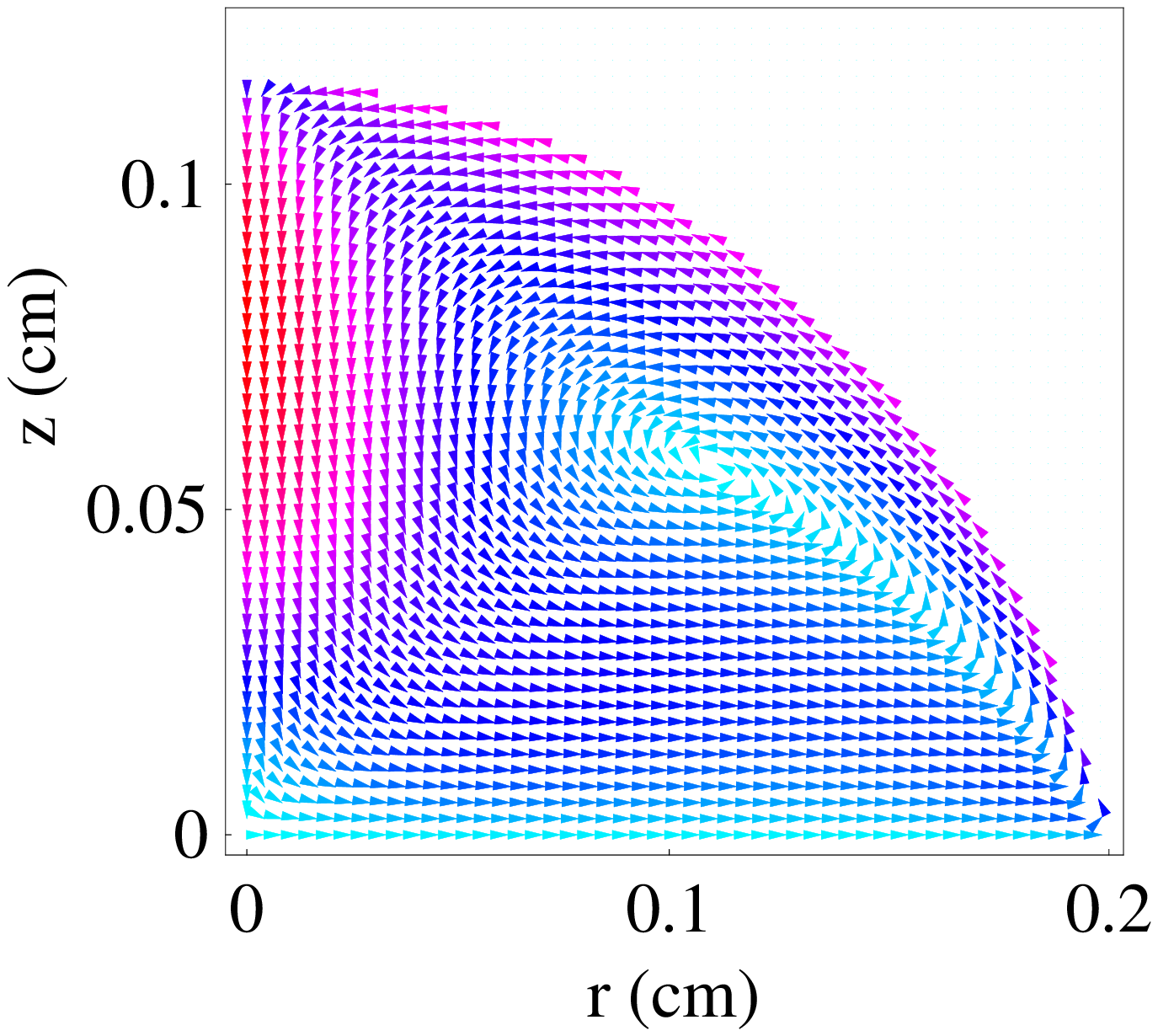}
\label{vect1}
\end{figure*}

\begin{figure*}[htb]
\caption{(Color online.) Distributions of temperature within drop for
the stages of drop dynamics with three vortices (left) and a single vortex
(right). The distributions are taken at $t=0.5$ s and at $t=30$ s
correspondingly. Temperature scale shown at the right column.} \label{temp1}
\hspace{0.7cm}
\includegraphics[width=0.7\columnwidth,clip=true]{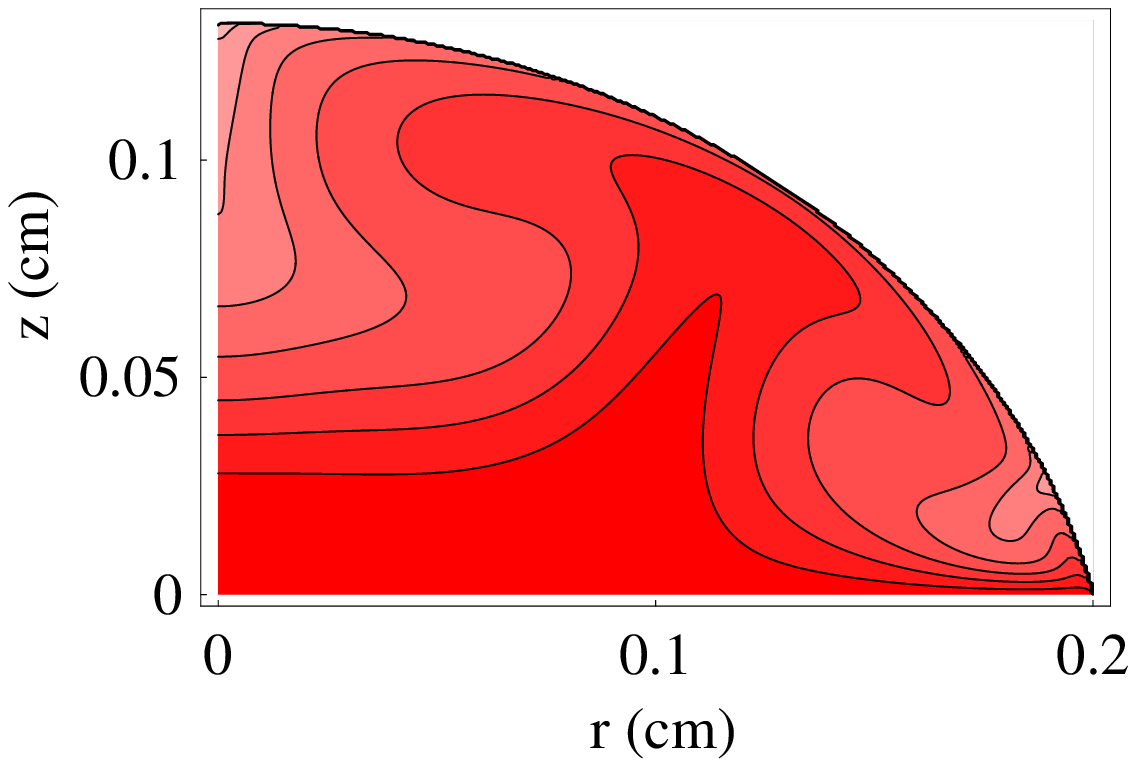}
\hspace{2.0cm}
\includegraphics[width=1.2\columnwidth,clip=true]{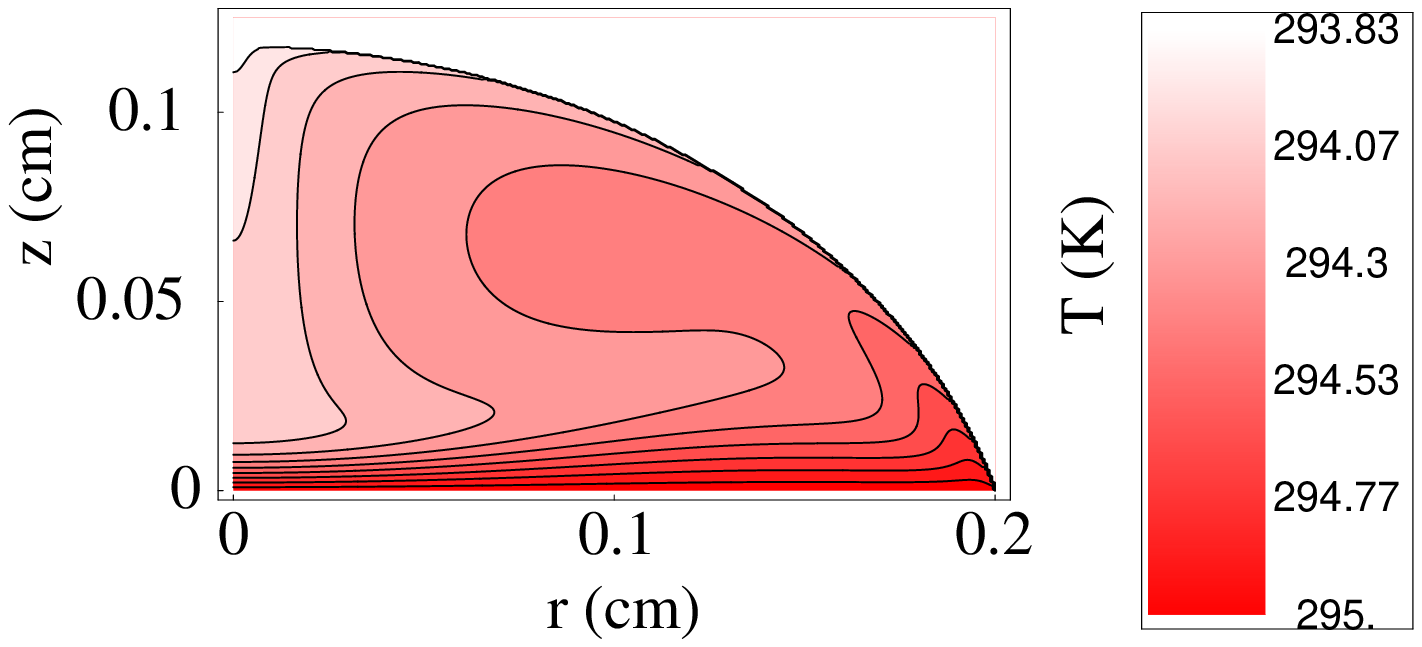}
\end{figure*}

During the enlargement of the near-surface vortices,
their number decreases, and the convection involves
the bulk of the drop. As a result, for $t\approx 0.45$ s,
three bulk vortices control the velocity and temperature
fields in the drop, as seen in left panels in
Figs.~\ref{vect2},\ref{temp1}.
During the coexistence of three vortices, the corner vortex
starts growing at the expense of
the other two vortices, and eventually at $t\approx 2.0$ s
it occupies the whole drop volume.
A spatial dependence of the temperature along
the drop surface is nonmonotonic, if the
drop contains more than one vortex (see Fig.~\ref{temp1}).
Right panels in Figs.~\ref{vect1},\ref{temp1}
demonstrate how in the single-vortex regime
effects of Marangoni forces drive liquid
along the surface to the apex, where the fluid penetrates along
the symmetry axis in the depth of the drop.

The regime with the single vortex represents one of the
main stages of the dynamics of the evaporating sessile drop.
It lasts up to $t\approx 250$~s.
More than half of the drop mass evaporates during this period of time.
If the initial values of mass, height and contact angle of the drop are
$m=8.7$~mg, $h=0.1314$~cm, $\theta=1.2045$, then at the moment $t=250$~s
we find $m=4.0$~mg, $h=0.0685$~cm, $\theta=0.716$. In particular,
$h/(2r_0)\approx 0.17$, i.e. the drop shape is noticeably flattened.
We remind that the total time of the evaporation of the toluene drop is $508$ s.

The quasistationary single-vortex state loses its stability at
$t\approx 250$ s and the vortex acquires a pronounced nonstationary
character. During this nonstationary regime, the fluid pulsations
take place. The characteristic frequency of the pulsations
corresponds to the circulation period $0.15$ s of a fluid element in
the original vortex. Initially the pulsations are concentrated near
the center of the original vortex. Then,
the single-cell pulsating
state breaks into two-center (and later three-center) pulsating
structure (see Fig.~10 in~\cite{Barash1}).
Eventually at $t\approx 300$ s a quasistationary state with
three vortices arises.


The numerical calculations of the fluid dynamics were tested with
several different mesh sizes. The respective results are
qualitatively identical and show reliable convergence
of the quantitative characteristics. For example,
the single-vortex regime was found to arise at $3.48$ s, $2.48$ s,
$2.2$ s, $2.06$ s, $2.01$ s for 100$\times$100, 150$\times$150,
200$\times$200, 250$\times$250, 300$\times$300 mesh elements
covering a half of the drop cross-section.

\section{Numerical method}

The simultaneous calculation of the physical quantities in the drop can
be partitioned into several steps:

\begin{enumerate}
\item We apply to the diffusion equation ${\p u}/{\p t}=D\Delta u$
the implicit finite difference method using irregular mesh
outside the drop and a variable time step. We use a boundary
interpolation in a vicinity of the drop surface~\cite{Barash1}.
For the boundary conditions we take $u=u_s$ on the drop surface,
$u=0$ far away from the drop, $\p u/\p r=0$ and $\p u/\p z=0$
on the axes $r=0$ and $z=0$ correspondingly.

In order to calculate accurately the vapor density in quite a large region,
it is convenient to use irregular mesh.
We use the mesh with sufficiently small steps near
the drop surface and with exponentially increasing
steps, which are chosen in accordance with the
decay of the vapor density, which can be estimated analytically.

\item Calculations of the stream function $\psi$ and
velocities $\vvect$ inside the drop are based on
Eq. ${\p^2\psi}/{\p r^2}-{\p\psi}/{r\p r}+
{\p^2\psi}/{\p z^2}=r\gamma$, where $\gamma$ is vorticity.
The implicit finite difference method with a regular mesh
inside the drop is applied. We use a boundary interpolation
near the drop surface.
For the boundary conditions we take $\psi=0$ at all boundaries:
on the surface of the drop and on the axes $r=0$ and $z=0$.

\item We solve Eq. ${\p\gamma(r,z)}/{\p t}+(\vvect\cdot\mynabla)\gamma(r,z)=
\nu \left(\Delta\gamma(r,z)-{\gamma(r,z)}/{r^2}\right)$ to obtain
the vorticity $\gamma$ inside the drop.
The explicit finite difference method with a regular mesh
inside the drop is used. We use a boundary interpolation
close to the drop surface.
For the boundary conditions we take
$\gamma=0$ for $r=0$; $\gamma=\p v_r/\p z$ for $z=0$;
$\gamma={d\sigma}/(\eta ds)+2v_\tau d\phi/ds$
on the drop surface, where ${d\sigma}/{ds}=-\sigma'\p T/\p s$ is
the derivative of the surface tension along the drop surface.

\item For calculating the temperature $T$ inside the drop,
the explicit finite difference method with a regular mesh
is applied to the thermal conduction equation
${\p T}/{\p t}+\vvect\cdot\mynabla T = \kappa \Delta T$.
We use a boundary interpolation in a vicinity of the drop surface.
The boundary conditions take the form $\p T/\p r=0$ for $r=0$;
$T=T_0$ for $z=0$; ${\p T}/{\p n}=-{Q_0(r)}/{k}=-LJ(r)/k$
on the drop surface.
Here $Q_0(r)$ is the rate of heat loss per unit area of the free
surface, $\mathbf{n}$ is a normal vector to the drop surface.

\item During the iterative procedure, the drop shape is recalculated
in accordance with the evaporative mass loss
for the respective time interval. This is possible due to a quasistationary character
of the shape change of the drop.
Deviations of a drop profile
from a spherical cap are taken into account~\cite{Barash1,Barash2}.

\end{enumerate}

\section{Discussion}

We have developed the approach for studying
the evaporation and fluid dynamics of a sessile drop of a capillary size
and applied it for the description of the toluene drop evaporation.
In particular, the approach allows to obtain the time evolution of 
vortex structure and temperature distribution in evaporating sessile drops.
The appearence of near-surface vortices on the early stage of the evaporation
process of sessile drops is predicted. 
We also obtained three bulk vortices in the intermediate stage.
They finally evolve into the single convection vortex
in the drop, existing during about $1/2$ of the evaporation time.
Possible development of the method include description
of hydrodynamics of colloidal drops, and study of influence of substrate conductivity
on the dynamics of vortex structure.

\end{document}